\documentclass[preprint,12pt,review,3p,sort&compress]{elsarticle}
\usepackage{graphicx}
\usepackage{dcolumn}
\usepackage{bm}
\usepackage{array}
\usepackage{float}

\newcolumntype{C}[1]{>{\centering}m{#1}}
\usepackage{amsmath} 
\usepackage{fancyhdr}
\usepackage{textcomp}

\usepackage{hyperref}
\hypersetup{
colorlinks=true,        
linkcolor=blue,          
citecolor=blue,         
urlcolor=blue            
}

\journal{Carbon}

\begin{document}

\begin{frontmatter}

\title{Surface passivation by graphene in the lubrication of iron: A
comparison with bronze}

\author[label1,label2,label3]{Diego Marchetto\corref{cor1}}
\address[label1]{Dipartimento di Scienze Fisiche, Informatiche e Matematiche, Universit\`a di Modena e Reggio Emilia, Via Campi 213/A, 41125 Modena, Italy}
\address[label2]{CNR-Institute of Nanoscience, S3 Center, Via Campi 213/A, 41125 Modena, Italy}
\address[label3]{Centro Interdipartimentale per la Ricerca Applicata e i Servizi nel Settore della Meccanica Avanzata e della Motoristica, Universita di Modena e Reggio Emilia, Via Vivarelli 2, 41125 Modena, Italy}
\cortext[cor1]{Corresponding author}
\ead{diego.marchetto@unimore.it}

\author[label1]{Paolo Restuccia}

\author[label1,label3]{Antonio Ballestrazzi}

\author[label1,label2]{M. C. Righi}

\author[label3]{Alberto Rota}

\author[label1,label2,label3]{Sergio Valeri}

\begin{abstract}
It has been recently reported that graphene is able to significantly reduce the friction coefficient of
steel-on-steel sliding contacts. The microscopic origin of this behavior has been attributed to the mechanical
action of load carrying capacity. However, a recent work highlighted the importance of the chemical
action of graphene. According to this work graphene reduces the adhesion of iron interfaces by reducing
the surface energy thanks to a passivation effect. The aim of the present work is to clarify the still
debated lubricating behavior of graphene flakes. We perform pin-on-disc experiments using liquid
dispersed graphene solution as a lubricant. Two different materials, pure iron and bronze are tested
against 100Cr6 steel. Raman spectroscopy is used to analyze the surfaces after the friction tests. The
results of these tests prove that graphene flakes have a beneficial effect on the friction coefficient. At the
same time they show a tendency of graphene to passivate the native iron surfaces that are exposed
during sliding as a consequence of wear.
\end{abstract}

\end{frontmatter}

\section{Introduction}

The recent need for friction and wear control in mechanical
high-technology applications has taken to the introduction of
layered materials (MoS$_2$, HBN, graphite) in coatings, additives,
liquid and solid lubricants~\cite{Scharf2013,Donnet2004,Levita2014,Wan2014,Podgornik2015,Bares2009,CutEdgTri}.
As a further scientific and technological improvement the potential of graphene as a lubricant is
currently under investigation. The reduction of friction obtained
with the application of a layer of graphene is studied at different
scales with different methods both theoretically~\cite{FricWear,Guo2007,Bonelli2009,Xu2011,Xu2012,Liu2011,
Smolyanitsky2012,Reguzzoni2012,Reguzzoni2012b,Hod2012,Leven2013,Cahangirov2013,Klemenz2014,Cahangirov2015} and
experimentally~\cite{Kim2011,Filleter2009,Lee2010,Lee2009,li2010,Lin2011,Marchetto2012,Marchetto2015,
Kandanur2012,Berman2014,Berman2014b,Ota2015,Liang2016,Lin2011b,Huang2006,Fan2015,Pu2014,Mao2015,
Feng2013,Deng2012,Berman2013,Berman2013b,Shin2011,Wahlisch2013}.

While the majority of the research at the nanoscale investigates
the properties of graphene as a solid atomic-thick coating
~\cite{Kim2011,Filleter2009,Lee2010,Lee2009,li2010,Lin2011,Marchetto2012,Marchetto2015}
at the macroscale graphene is studied also as colloidal liquid
lubricant~\cite{Kandanur2012,Berman2014,Berman2014b,Ota2015,Liang2016,Lin2011b,Huang2006,Fan2015}.
In few recent works, Berman et al. successfully
used graphene to lubricate a steel-steel sliding contact~\cite{Berman2014,Berman2014b,Berman2013,Berman2013b}.
Graphene flakes dispersed in an alcohol solution
were applied to a steel surface and tested by tribometer. In humid
environment the reduction of friction was dramatic~\cite{Berman2013,Berman2013b}. While
the steel on steel coefficient of friction is usually around 0.9 in the
presence of graphene it dropped to about 0.2~\cite{Berman2013}.
In inert environment the same values for the friction coefficient were found
together with a marked reduction of wear~\cite{Berman2013,Berman2013b}. The lifetime of
the graphene layer and therefore of the lubricated regime
decreased with the interface pressure~\cite{Berman2013b}. The short lifetime of
graphene layers at high loads has been observed in other works~\cite{Klemenz2014,Marchetto2015}.

A theoretical work on the processes governing the tribology of
metal-supported graphene was published in 2014 by Klemenz
et al.~\cite{Klemenz2014}. They combined atomistic simulation of nanoindentation
and AFM experiments on graphene covered Pt(111) surfaces. The
reduction of friction coefficient achieved with the application of
graphene was explained by the authors with the ability of
graphene to increase the load carrying capacity of the surface~\cite{Klemenz2014}.
However, when graphene gets damaged it loses its lubricating properties~\cite{Klemenz2014}.

Restuccia et al. instead propose that the lubricating behavior of
graphene is due to its ability to passivate the steel surfaces~\cite{Restuccia2016}.
They show that the carbon dangling bonds of graphene chemically
adsorbs on native iron surfaces produced during sliding screening
the metal-metal interaction. Due to this behavior the atomic-thick
carbon layer is able to reduce the interfacial adhesion and shear
strength~\cite{Restuccia2016}.

Therefore while the work of Klemenz et al. takes into account
mechanical lubrication~\cite{Klemenz2014}, the one of Restuccia et al. involves
tribochemical lubrication~\cite{Restuccia2016}.

With this paper we add a new piece of information for understanding the lubricating properties of graphene by analyzing the
role played by the surface chemistry from an experimental point of
view. We perform ball-on-disc friction tests on pure iron and on
bronze using a solution of graphene flakes in ethanol. Performing
Raman analysis on the samples (discs and balls) before and after the
tests we analyze the status of the graphene flakes and identify their
main function. It must be said that the case described by Klemenz
et al. is different from the one analyzed here. They studied
epitaxially grown graphene without any defects while we study the
behavior of deposited graphene flakes.

\section{Materials and methods}

The tribological tests are performed by means of a commercial
tribometer CETR UMT-3 in ball-on-disc configuration similarly to
previous studies on liquid dispersed solution of graphene~\cite{Berman2014,
Berman2014b,Fan2015,Berman2013,Berman2013b}. The ball used for all tests is
4 mm in diameter made out of steel 100Cr6.

The tested samples are made of iron (99.98\% pure) and bronze
(98\% Cu and 2\% Sn). The discs are all polished in order to obtain a
30 nm surface average roughness $R_a$.

Tests are performed maintaining a tangential sliding velocity of
100 mm/s (at a radius of 10 mm) and the applied load is 1 N. This
load produces a nominal average pressure of about 0.4 GPa (Hertzian pressure).
All tests are made at room temperature with a $RH$ humidity of about 50\%.

\begin{figure}[htpb]
 \begin{center}
\includegraphics[width=\linewidth]{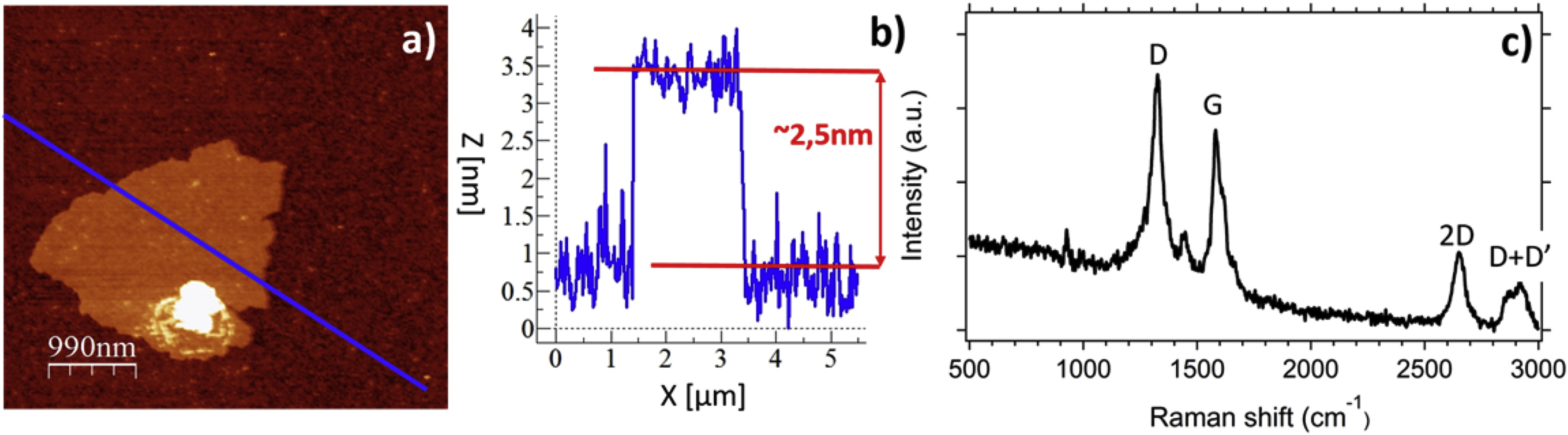}
  \caption{a) AFM picture of a flake of graphene deposited on SiO$_2$. The image shows that some kind of contamination is present. The lateral dimension correspond to the specifics. b) Profile. The thickness is quite high and corresponds to about three layer of graphene. c) Raman spectrum of graphene deposited on a clean iron surface. The characteristic peaks D, G and 2D (at 1330, 1590 and 2650 cm$^{-1}$) of graphene are observed. The intensity of the D peak proves that the flakes present a large amount of defects, edges or partial oxidation.}\label{fig:Fig1}
\end{center}
\end{figure}

For the tests with graphene we used a commercial liquid
dispersed solution of graphene single layers (Graphene Supermarket).
The dispersion is in ethanol with a concentration of the
graphene flakes of 1 mg/L. The flakes should not present either
oxidation or surfactants. The average flake size is 550 nm
(150-3000) nm. Tests performed by AFM on some flakes deposited
on SiO$_2$ show that the flakes are actually multilayered and present
defects and contamination (see Fig.~\ref{fig:Fig1}a and b). In Fig.~\ref{fig:Fig1}c
the Raman spectrum of graphene flakes deposited on a clean iron surface is
shown. The D, G and 2D peaks of graphene are present. The intensity of the D peak
(related to the breathing mode of the sp2 atoms) indicates that the amount of crystalline
defects, edges or oxidation is rather strong for such commercial flakes, and probably
fluctuating among different batches (see the qualitatively different
spectrum in Fig. 1b of \cite{Berman2013}, for the very same graphene solution).
The spectrum also confirms that the flakes are not single layer but
more probably made of two or three layers. This is compatible with
previous studies~\cite{Fan2015,Ferrari2006}.

Drops of graphene solution are applied directly on the wear
track during the friction tests. Each drop is about 0.05 ml. Before the
tests some drops are applied and let exsiccate on the surface both of
Fe samples and of bronze ones. This procedure already tested by
Berman et al. produces a layer of graphene flakes on the sample
surfaces~\cite{Berman2013,Berman2013b}.

The surfaces are analyzed by Raman spectroscopy [He-Ne Laser,
632.81 nm e Jobin Yvon] inside and outside the wear-tracks before
and after the tests.

\section{Experimental results}

\begin{figure}[htpb]
 \begin{center}
\includegraphics[width=\linewidth]{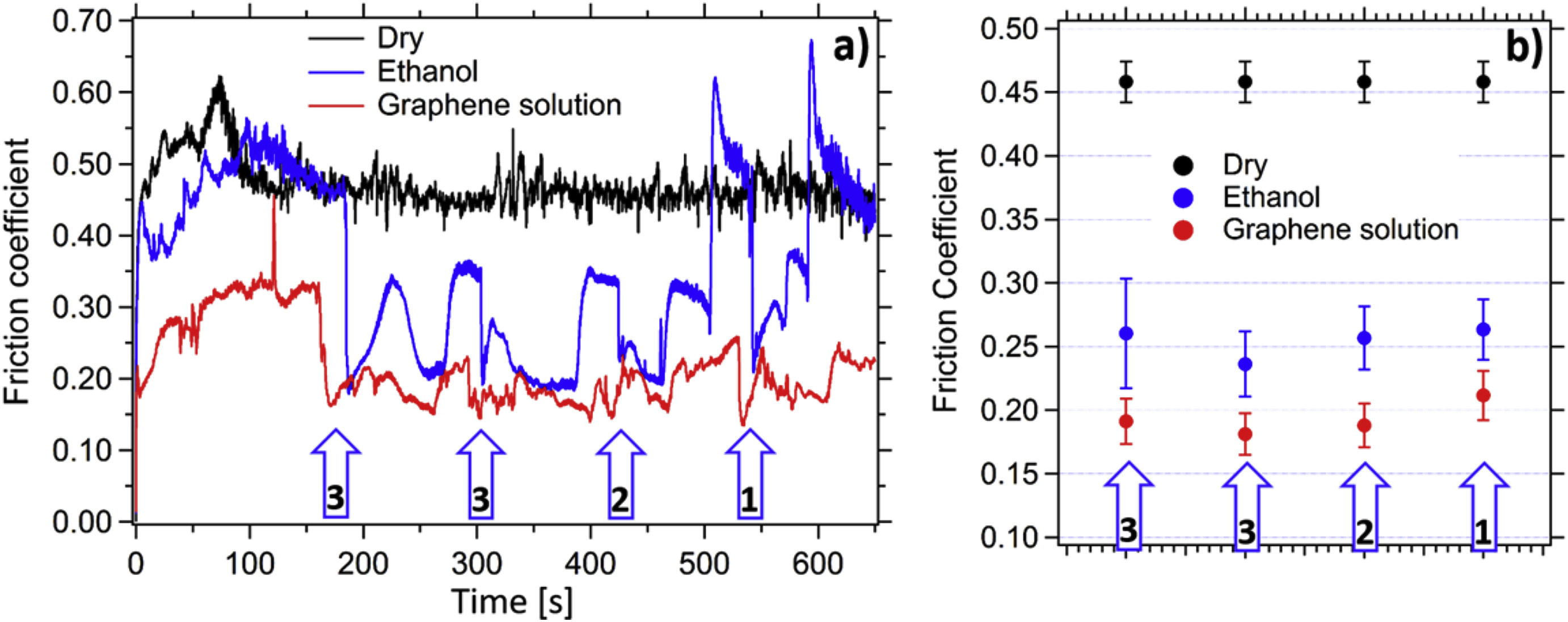}
  \caption{Friction coefficient results. a) example of the friction coefficient behavior in a dry contact (black), in a contact lubricated by ethanol (blue) and in a contact lubricated by ethanol solution containing graphene (red). The arrows represent the application time when droplets are poured, the number of which is indicated by the arrow; b) Average values of the friction coefficient obtained averaging the data of the tests after the solution application.}\label{fig:Fig2}
\end{center}
\end{figure}

An example of friction results is reported in Fig.~\ref{fig:Fig2}a. The graph
shows a comparison between tests performed in ``dry'' conditions,
a test performed adding drops of graphene solution and a test performed
adding drops of pure ethanol. Drops are added in both cases
every two minutes in the amount of three at the first step, three at
second, two at the third and one at the fourth step. In dry conditions the
friction coefficient after the running-in becomes stable
around a value of about 0.46. The test performed adding graphene
show a completely different behavior. The COF starts at a value of
0.05 and during the running-in phase rises to 0.32. This is due to the
presence of the graphene layer produced before the test. After three
minutes the first three drops are applied and the COF goes down to
0.15. Although with some oscillations the COF does not move from
this value for the rest of the test and drop steps.

Using ethanol drops (blue line Fig.~\ref{fig:Fig2}a) the behavior is different.
The friction reduction after applying few drops of ethanol is
evident, but the coefficient rises as soon as the ethanol evaporates.
When just one drop is poured (last step) the evaporation time is
quite short and the lubrication process lasts for few cycles only.

In Fig.~\ref{fig:Fig2}b the average values of the friction coefficient are reported.
They are calculated by averaging the data recorded between
the pouring moments. For ethanol the data obtained after the
evaporation of the liquid were excluded from the average. In this
way it is possible to compare the lubrication effect of liquid ethanol
with that of ethanol containing graphene flakes. As can be seen in
Fig.~\ref{fig:Fig2}a, the friction coefficient in dry conditions is about 0.46.
Introducing ethanol we obtain a $\mu = 0.24$ with some fluctuations.
Graphene flakes lower the friction coefficient to a value of $\mu = 0.19$.
In few occasions the pure ethanol and the graphene-containing
solution have comparable values (mainly due to big fluctuations
in the ethanol $\mu$ values) but considering the whole picture it is clear
that graphene not only improves the duration of the low friction
regime, but also improves the lubricating effect in terms of absolute
value as well.

\begin{figure}[htpb]
 \begin{center}
\includegraphics[width=\linewidth]{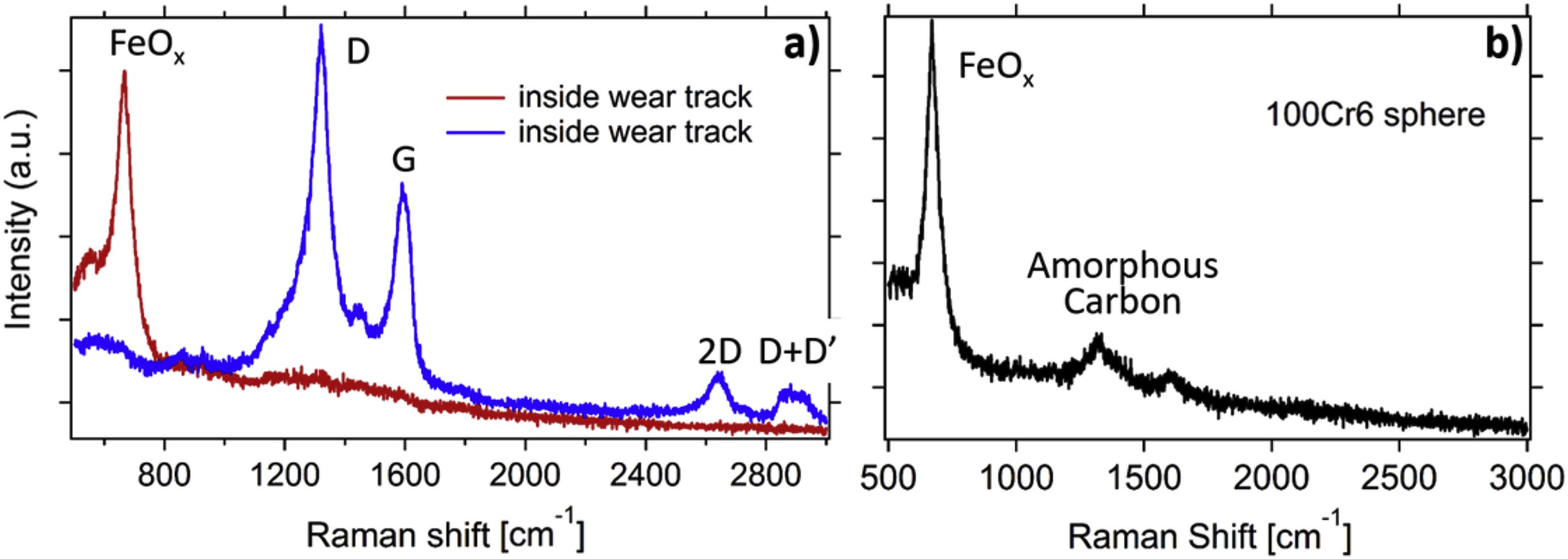}
  \caption{a) Raman spectroscopy of a wear track on Fe samples. The friction tests was stopped few minutes after pouring graphene drops. Ethanol was completely evaporated and the friction coefficient was still low around $\mu = 0.16$. The red curve show no carbon at all, while the blue curve shows a clear presence of graphene. b) Raman spectroscopy of the sphere
used for the test. The shape of the curve shows the presence of iron oxide and amorphous carbon.}\label{fig:Fig3}
\end{center}
\end{figure}

The Raman analysis performed on the samples is reported in
Fig.~\ref{fig:Fig3}. Since the friction test is carried out after the solution is
completely evaporated from the sample and the friction coefficient
is still low (at a value of about 0.16) it is possible to investigate the
presence of graphene both on the wear-tracks and on the ball. The
results reported in Fig.~\ref{fig:Fig3}a show that two different kinds of spectra
can be found within the wear track. One spectrum corresponds to
iron oxide. No graphene is visible. This can be associated with the
oxidation of the surface and the formation of wear debris due to
sliding. The second spectrum clearly shows the presence of graphene.
The 2D peak at 2646 cm$^{-1}$ is clearly visible and so are D and
G peaks proving that the layer is not single and that defects are
definitely present. The 2D peak is smaller than the one measured
on a pristine layer (Fig.~\ref{fig:Fig1}) while D and G are much more intense.
The D peak in particular reveals that the flakes of graphene are
highly damaged. This should not be a surprise considering that
those flakes were laying in between the two sliding surfaces
standing a pressure of several hundred MPa. A similar situation is
shown by Mao et al. for flakes applied in electric contacts~\cite{Mao2015}.

In Fig.~\ref{fig:Fig3}b the Raman spectrum from the ball shows that on the
sphere in the contact area there is actually no graphene. D and G
peaks are barely visible and the 2D peak is not visible at all, as it
happens in amorphous carbon~\cite{Mao2015}. So on the ball there is just a
transfer of carbon residues from the grinding of graphene during sliding.

\begin{figure}[htpb]
 \begin{center}
\includegraphics[width=\linewidth]{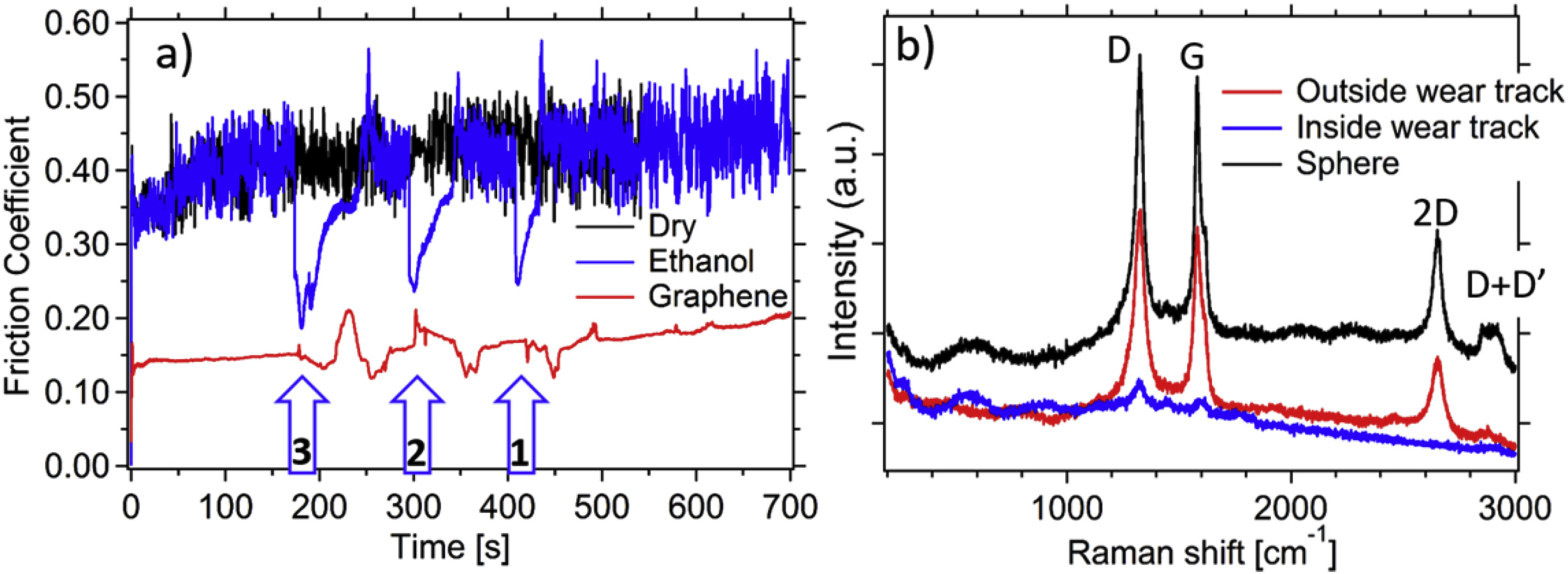}
  \caption{a) Pin-on-disc results obtained on a bronze disc. The arrows represent the number of drops (indicated on the arrow) and the instant at which they are poured. The black line represents the test in dry conditions which is shorter; b) Raman spectra obtained on the bronze sample and on the ball.}\label{fig:Fig4}
\end{center}
\end{figure}

The ball-on-disc tests are repeated on a bronze disc. The same
set-up described above is used. The results are reported in
Fig.~\ref{fig:Fig4}a. The ethanol on bronze gives almost none lubrication. We
observe a drop in the friction coefficient but the intensity and
duration of it is very much different from those observed on iron.
After a few cycles of lubrication the friction coefficient rises back
to the ``dry'' values. The system lubricated by the graphene solution shows a stable value of
$\mu = 0.15$ from the very beginning. Only after putting the first three drops some
instabilities in the friction appear.

The Raman spectra obtained on the bronze sample are reported
in Fig.~\ref{fig:Fig4}b. The wear-track corresponds to a test stopped in the low
friction regime after drops of the graphene-containing solution
have been poured and the ethanol is evaporated. The curves show
that graphene is present outside the track (red curve) but not inside
(blue curve). Measurements performed on the ball inside the
contact area show very intense peaks of graphene (black curve).
The graphene D peak is quite high (as well as on iron sample)
indicating that the graphene flakes attached to the ball contain
defects produced by the friction process.

\section{Discussion}

The results reported above give us a quite clear picture of the
behavior of the graphene flakes. Starting from Fig.~\ref{fig:Fig3}, the presence of
graphene in the wear track verified by the Raman spectroscopy
confirms that the flakes have a lubricating effect. The Raman
analysis performed on wear tracks after the friction coefficient is
back to ``dry values'' ($\mu = 0.45$) do not show any presence of graphene.
These data are close to the results presented in literature by
Berman et al.~\cite{Berman2014,Berman2014b,Berman2013,Berman2013b}.
On the other hand the duration of the lubricating effect is totally different.
Berman et al. report thousands of cycles at low friction coefficient while in our
experiment we reach a few hundred at best (once we stop pouring solution drops).
This difference is ascribable to the quality of graphene. The flakes
used for this experiment are highly defected and partially oxidized
while the graphene used by Berman et al. is of a much higher
quality. This statement is confirmed by the comparison of the
Raman spectrum reported in Fig.~\ref{fig:Fig1}c and the one reported by Berman~\cite{Berman2013b}.

The presence of completely uncovered areas in the wear track
on which the Raman spectrum shows only the iron oxide peak,
reveals that the coverage is not complete. Furthermore the Raman
spectrum of the graphene flakes presents a D peak that is double
the size of the G one. This is due to a presence of defects. Finally the
Raman spectrum of the sphere does not show any presence of
graphene. This last result is a bit surprising. We would indeed
expect graphene to cover both surfaces.

The picture becomes much clearer when the results of Fig.~\ref{fig:Fig3} are
compared to the results of Fig.~\ref{fig:Fig4}. On bronze the behavior of graphene is
completely different. No carbon material is found in the
wear track although the friction coefficient induced by graphene
lubrication is even more stable than the one on Fe. Graphene on the
sample is found only outside the wear track as expected (from the
layer are produced before the tests). Instead a very clear signature
of graphene is visible on the 100Cr6 sphere inside the area of
contact. The intensity of the peaks D, G and 2D is comparable to that
of the peaks produced by the initial layer on the bronze sample
(Fig.~\ref{fig:Fig4}). This fact implies that the graphene on the ball is not as
damaged as the graphene found in the wear-track on the Fe sample.

The fact that graphene is found on the sphere and not on bronze
indicates that graphene flakes adsorb most preferably on iron than
on bronze. This experimental observation can be explained by
comparing the graphene adsorption energy on iron and copper (the
bronze sample used in the experiment contains 98\% copper). The
binding energy per carbon atom obtained by DFT calculations
within the local density approximation (LDA) corresponds to
36 meV on copper and to 134 on iron (full calculations are shown in
Supplementary materials). The comparison of the binding energies
and the equilibrium distances, 3.2 \r{A} on copper and 2.04 \r{A} on iron,
indicates that there is a marked difference in the nature of the interactions:
graphene physisorbs on copper, while it chemisorbs on iron, where the pi orbital 
of graphene hybridize with the d states of iron that are not fully occupied.
The different nature of the interactions, is also reflected in the electronic
structure of adsorbed graphene, which preserve its main feature on copper, while it is
strongly perturbed on iron~\cite{Khomyakov2009,Vinogradov2012}.

Graphene does not cover the entire surface but work as an oil
additive. During sliding of the steel sphere against the discs (both
bronze and Fe) some wear occurs. In the case of pure iron the
majority of the wear is located on the surface of the disc. Fe is softer
than steel and the superficial iron oxide layer is continuously worn
out.

This process uncovers native Fe surfaces, which are chemically
very active. As explained by Restuccia et al.~\cite{Restuccia2016} the graphene
flakes dispersed in the solution stick to this spots reducing the
surface energy and therefore the friction coefficient. This behavior
explains the reduction of friction and the instability of the friction
coefficient. Indeed, the coating is not complete and once the
ethanol is completely evaporated the flakes cannot move
anymore. Therefore the sphere asperities encounter areas with
low friction due to the presence of graphene covering Fe and areas
at higher friction with uncovered Fe bonds or iron oxide. The fact
that the Raman spectra obtained from these tests show very
damaged graphene could be due both by the presence of an oxide
layer on iron that induces oxidation also in graphene during
sliding and by the defects originating from the high pressure at
the asperity-asperity contacts.

On the bronze sample the flakes do not stick to the wear track.
Raman spectrum only shows the presence of some amorphous
carbon due to the presence of carbon material (coming from graphene, the environment,
and ethanol). On the other hand the 100Cr6 steel sphere is well covered by graphene.
The flakes bind to the Fe bonds on the ball surface (100Cr6 is about 97\% iron). The
better stability and duration of the low friction regime obtained on
bronze compared to steel is probably due to the higher ``graphene
flakes concentration/Fe active bonds'' ratio. The graphene coverage
is higher and therefore the surface energy is lower. This result is in
good agreement with the outcome of the analysis on the effects of
graphene coverage performed in Ref.~\citenum{Restuccia2016} by means of first principles calculations.
As a further proof of this behavior, an analysis of
some of the wear tracks (reported in Supplementary materials)
shows that the wear reduction induced by the graphene lubrication
in bronze tests is higher than that observed in tests performed on
iron.

\section{Conclusions}

We performed tribological experiments and spectroscopic
analysis to shed light into the chemical mechanisms underlying
lubricating effect of graphene. An ethanol solution containing
graphene flakes was used to lubricate a steel-iron and a steel-bronze
interface in ball-on-disc friction tests. The reduction of
friction coefficient due to graphene was compared to the one obtained with
the use of pure ethanol and to the ``dry'' condition. A
reduction of friction ascribable to graphene was measured both in
terms of absolute value and in terms of duration compared to the
pure ethanol.

The Raman spectroscopy performed on the disc surfaces and on
the contact area of the balls showed that graphene flakes binds to
iron rather than other materials. Indeed in the steel-iron sliding
contact graphene was found on the Fe surface while in the steel-bronze
system graphene bound to the active region of native iron
which were produced on the steel ball by rubbing. We were
therefore able to show that the lubrication due to the graphene
flakes is related to the chemical passivation of iron by graphene.

The difference between the friction coefficients obtained on
pure iron and on bronze were attributed to the different coverage of
the surfaces. In the case of steel ball sliding on bronze the small
contact area of the sphere is quickly covered by graphene flakes
inducing a more stable and durable lubricating effect.

\section*{Acknowledgments}

The authors gratefully acknowledge the support by Cost Action
MP1303 ``Understanding and controlling nano and mesoscale friction'
and FCRMo for support through the project SIME 1013.0650
(call ``Applied Research for Innovation''). We thank Renato Buzio
from CNR-SPIN for useful discussion.

\appendix
\section{Supplementary data}

Supplementary data related to this article can be found at
\href{http://dx.doi.org/10.1016/j.carbon.2017.02.011}{link}.

\section*{References}

\bibliography{biblio}
\bibliographystyle{elsarticle-num}

\end{document}